\pdfoutput=1
\documentclass{article}
\usepackage{amsmath}
\usepackage{algorithm}
\usepackage{algpseudocode}
\usepackage{physics}
\usepackage{indentfirst}
\usepackage{multirow}
\usepackage[section]{placeins}
\usepackage{float}
\usepackage{authblk}

\usepackage[english]{babel}

\usepackage[letterpaper,top=2cm,bottom=2cm,left=3cm,right=3cm,marginparwidth=1.75cm]{geometry}

\usepackage{amsmath}
\usepackage{graphicx}
\usepackage{subcaption}
\usepackage[colorlinks=true, allcolors=blue]{hyperref}

\title{Low-Rank Expectile Representations of a Data Matrix, with Application to Diurnal Heart Rates} 
\author[1]{Shuge Ouyang\textdagger}
\author[2]{Yunxuan Tang\textdagger}
\author[1]{Benjamin Osafo Agyare\thanks{Advised by Kerby Shedden.}}

\affil[1]{Department of Statistics, University of Michigan}
\affil[2]{Department of Computer Science Engineering, University of Michigan}
\date{}
\begin{document}
\maketitle
\renewcommand{\thefootnote}{\fnsymbol{footnote}}
\footnotetext[1]{\textdagger~Equal contribution.}

\renewcommand{\thefootnote}{\arabic{footnote}}

\begin{abstract}
    Low-rank matrix factorization is a powerful tool for understanding the structure of 2-way data, and is usually accomplished by minimizing a sum of squares criterion. Expectile analysis generalizes squared-error loss by introducing asymmetry, allowing tail behavior to be elicited. Here we present a framework for low-rank expectile analysis of a data matrix that incorporates both additive and multiplicative effects, utilizing expectile loss, and accommodating arbitrary patterns of missing data. The representation can be fit with gradient-descent. Simulation studies demonstrate the accuracy of the structure recovery. Using diurnal heart rate data indexed by person-days versus minutes within a day, we find divergent behavior for lower versus upper expectiles, with the lower expectiles being much more stable within subjects across days, while the upper expectiles are much more variable, even within subjects.
\end{abstract}

\section{Introduction}




Matrix models have been widely employed for understanding data indexed by two variables. In this work, we represent our matrix model as $X=R1^\top+1 C^{\top}+U V^{\top}+E$. In this formula, $R$  and $C$ are vectors of row and column effects, representing the heterogeneity in the data. $U$ and $V$ are matrices of latent factors, where the outer product $U V^{\top}$ captures the higher-order dependencies and clustering patterns. This matrix representation is based on Peter Hoff's additive and multiplicative effects model (AME) \cite{10.1214/19-STS757}. The AME model is capable of capturing various types of statistical dependencies in network data, including node-level correlations, node heterogeneity, and higher-order dependencies such as transitivity and clustering. We apply the AME model to analyze heart rate data with missing values, which comes from a study conducted at RTI International Research Institute
(RTI) in 2016 \cite{furberg_2016_53894} capturing 24-hour (circadian) heart rates in 14 human subjects over the course of multiple weeks, obtained using wearable sensors.

In this model, a key contribution is that expectiles are introduced in the setting of matrix factorization. Expectiles are summary statistics that characterize a probability distribution $F$ over a continuum of location parameters, indexed by a value denoted $\tau$ ($0 < \tau < 1$).  Expectiles result from an asymmetrically weighted version of ordinary least squares in that the expectile equals the mean when $\tau=1/2$, and in general the expectile minimizes a weighted squared error loss where the positive residuals are weighted by $\displaystyle\frac{\tau}{1-\tau}$ and the negative residuals are assigned unit weights. For a random variable $X$ with distribution $F$ and a finite first moment, $E|X|<\infty$, Newey and Powell \cite{newey1987asymmetric} introduced executives as the set of minimizers:
$$
\mu_\tau:=\arg \min _\mu \int \varsigma_\tau(x-\mu) d F(x),
$$
where the function $\varsigma_\tau$ is an asymmetric least squares criterion:
$$
\varsigma_\tau(u)= \begin{cases}\tau u^2 & \text { if } u \geq 0 \\ (1-\tau) u^2 & \text { otherwise. }\end{cases}
$$

Expectile regression uses the concept of expectiles to study conditional distributions of an outcome given predictors.  It is an innovative statistical method gaining traction in recent years, and offers a unique approach to analyzing distribution tails. The expectile of a distribution is related to but distinct from the more familiar concept of a quantile. Applications are largely from the field of econometrics, one case in point is in 2021 when Phillips \cite{philipps2021expectile} applied this method to Home Mortgage Disclosure Act data from Boston to find a link between mortgage application cupping and racial disparities. 

Low-rank representations of matrices using expectile loss provide a powerful strategy for handling the vast and complex datasets typical of human data. Low-rank representations are crucial for tasks such as dimensionality reduction, data compression, and noise reduction. They are particularly effective in matrix completion, collaborative filtering, and image processing, enhancing the identification and characterization of circadian patterns in heart rate data. This combination improves the robustness of the analysis against noise and missing data, facilitating the extraction of key features from heart rate data.


\section{Model Architecture}

Our model for expectile analysis includes both additive effects and low-rank multiplicative effects. This section discusses the theoretical derivation of the expectile loss function, its gradients, and optimization via gradient descent to obtain parameter estimates. 

\subsection{Derivation of the expectile loss function}

Define $X$ as the $n \times p$ observed data matrix, which may have missing values, and let $R$ $C$, $U$, and $V$ denote the model parameters. Specifically, $R$ is the $n \times 1$ vector of additive row effects, $C$ is the $p \times 1$ vector of additive column effects, $U$ is the $n \times k$ matrix of multiplicative row effects, and $V$ is the $p \times k$ matrix of multiplicative column effects. Then, we can define the following parameters for each non-missing entry:

The fitted low-rank matrix corresponding to $X$ is

\begin{center}
    $\widehat{X} = \widehat{R}_{n \times 1}1^{\top}_{p} + 1_{n}\widehat{C}_{p \times 1}^{\top} + \widehat{U}_{n \times k}\widehat{V}^{\top}_{p \times k}$,
\end{center}

\noindent and the residuals are

\begin{center}
    $\widehat{E} = X - \widehat{X} = X - (\widehat{R}_{n \times 1}1^{\top}_{p} + 1_{n}\widehat{C}_{p \times 1}^{\top} + \widehat{U}_{n \times k}\widehat{V}^{\top}_{p \times k})$.
\end{center}

\noindent We can thus define the weight parameter as $W_{ij} = W_{ij}(\widehat{E}) = \begin{cases}
    \tau & \widehat{E}_{ij} \geq 0\\
    (1 - \tau) & \widehat{E}_{ij} < 0.
\end{cases}$

\noindent To accommodate missing values, define the mask matrix $M$ to be the $n\times p$ matrix such that $M_{ij}$ is 1 if $X_{ij}$ is observed and $M_{ij}$ is zero otherwise. We can then define the loss function as

\begin{equation}
    L(R, C, U, V) = N^{-1}\sum_{i = 1}^{n}\sum_{j = 1}^{p} M_{ij} \cdot W_{ij}(\hat{E}) \cdot \widehat{E}_{ij}^2, \label{loss}
\end{equation}

\noindent whose first derivative (gradient) is defined everywhere as

\begin{eqnarray}
    \partial L / \partial R &=& -2N^{-1}(M \cdot W \cdot \widehat{E})1_{p}^\top\nonumber\\
    \partial L / \partial C &=& -2N^{-1}(M \cdot W \cdot \widehat{E})^\top1_{n}\nonumber\\
    \partial L / \partial U &=& -2N^{-1}(M \cdot W \cdot \widehat{E})V\nonumber\\
    \partial L / \partial V &=& -2N^{-1}(M \cdot W \cdot \widehat{E})^\top U.
\end{eqnarray}

We adopt the partial derivatives above in our loss minimization algorithm, and the complete derivation is included in Appendix~\ref{appendix.a}. It will be much more complicated when $W$'s partial derivatives concerning $R/C/U/V$ are undefined. We only evaluated and harnessed the calculation of the above gradient in our model. Studies in the future can expand on examining the gradient-descent of a low-rank model for expectile values further. 

\subsection{Data Generation and Preprocessing}

To compare the performance of different optimization algorithms and test the resilience of these algorithms against random initialization, we conducted some simulation studies with preprocessed data generated from the Gaussian distribution. 

We generated a simulation dataset drawn from the Gaussian Distribution with thirty percent missing data, which simulates the public dataset of heart rates we have access to. The simulation dataset is generated through an algorithm where we specify the number of rows and columns of the desired data, the standard deviation of both additive and multiplicative terms fitting a normal distribution, and the standard deviation of the true error to the entire dataset. We then randomly assign thirty percent of the data to be missing values to simulate the proportion of missingness in the actual dataset. 
After generating the simulated data, we also applied normalization to it so that the low-rank model could fit faster with normalized outputs, and made it easier for us to directly compare output results across trials. 
The algorithms for data generation and normalization are included in Appendix~\ref{appendix.b}.

\subsection{Model Fitting}
The next step was to fit our low-rank model with the normalized data. Since we calculated the row means and the column means of the normalized data, we initialized the additive terms $R$ and $C$ with corresponding row means and column means. This initialization reduces some randomness in the effect of $R$ and $C$, and hence focuses more on the initialization effect of $U$ and $V$, which is the main varying component of our model. To initialize $U$ and $V$, we employed the standard normal distribution consistently. 

The initialized parameters were then passed to \textbf{scipy.minimize} method, which updated expectile loss and gradients based on derivations in earlier sections. The underlying optimization algorithm was chosen from \textbf{BFGS, LBFGS,} and \textbf{CG}, whose final loss values and efficiency were evaluated in later sections. The detailed pseudo-code is accessible in Appendix~\ref{appendix.b}.

\subsection{Result Standardizing}
Finally, we standardize the matrix obtained from the decomposition to regularize our model and ensure that the resulting components are more consistent with different randomization of trials. To make it an identified mode, we enforced each column of multiplicative columns ($V$) has have a mean of 0, the additive column ($C$) to have a mean of 0, and normalized the multiplicative rows ($U$). We also used a method to ensure the orientation of the multiplicative row ($U$) is maintained throughout different initializations when the matrix rank is one. This greatly simplified the process of interpreting plots. We explain both procedures in Appendix~\ref{appendix.b}.

\section{Simulation Study}
We carried out a series of simulation studies utilizing the model architecture above to decide which optimization algorithm we would use and analyze our model performance regarding final loss values, mean absolute difference in the recovered data, and the effect of the chosen model rank, all of which paved the way for the analysis of the heart rate data.

\subsection{Simulate Data Configuration}
The heart rate dataset yields a matrix of dimensions $288 \times 334$, corresponding to 288 time bands and 334 person-days. To mirror this structure, our simulated dataset is crafted as a $200 \times 200$ matrix. In the simulation, the standard deviations for $R$, $C$, $U$, and $V$ are set to 1, and the means are set to 0. The simulation incorporates a missing data scheme, in which $30\%$ of the entries in the matrix are missing completely at random (MCAR). Furthermore, the matrices $U$ and $V$ possess a true rank of 2. Finally, additive independent Gaussian noise with residual standard deviation $\sigma = 0.3$ is incorporated.

\subsection{Accuracy of the fitted values}

For Mean Squared Error, we understand that the expectile loss value for $\tau = 0.5$ equates to the mean squared error of our prediction. This is given by the expression $\displaystyle\frac{\sigma_{\text{new}}^2}{2}$ \cite{newey1987asymmetric}, where $\sigma_{\text{new}}$ denotes the new error standard deviation after assigning missing values and recalculating the standard deviation. In our simulated study, we estimate $\sigma_{\text{new}} \approx 0.143$, leading to an expectile loss value of $\displaystyle\frac{\sigma_{\text{new}}^2}{2} \approx 0.0102$. Correspondingly, our model reports a loss value of $0.0098$ when $\tau = 0.5$, which corroborates our theoretical expectations with slightly overfitting. Consequently, this concordance suggests that the model achieves commendable loss values across all $\tau$ levels. 

\subsection{Performance of Loss Minimization Under Different Optimization Algorithms}

In our pursuit to minimize the loss function, we have selected three candidate gradient-based algorithms: Broyden–Fletcher–Goldfarb–Shanno algorithm (\textbf{BFGS}) \cite{35d0019d-775a-3628-b0b4-67be112e346b}, Limited-Memory Broyden–Fletcher– Goldfarb–Shanno algorithm (\textbf{LBFGS}) \cite{liu1989limited}, and Conjugate Gradient method (\textbf{CG}) \cite{hestenes1952methods}. To evaluate the performance and efficiency of these algorithms, we repetitively generated different simulated datasets as described above 100 times, and each time we conducted 100 initializations with specific $\tau = 0.2$ and rank $k = 3$. We recorded the minimum loss and time of each algorithm and found the algorithm with the smallest minimum loss and time respectively.

The aggregated results, depicted in the table \ref{tab:algorithm_simulation} below, for each of the 100 different simulated datasets, \textbf{CG} has the minimum final loss almost evenly, but the differences in loss values among the three algorithms are extremely small, less than $10^{-7}$. While \textbf{LBFGS} outperforms significantly in the time of execution. It's about 100 times faster than \textbf{BFGS} and 2-3 times faster than \textbf{CG}.

\begin{table}[H]
    \centering
    \begin{tabular}{|p{2cm}|p{4.5cm}|p{4.5cm}|}
    \hline
     \textbf{Algorithm}    & Number of times with minimum final loss & Number of times with minimum time of execution \\
     \hline
     BFGS    & 0 & 0\\
     \hline
     LBFGS & 0 & 100\\
     \hline
     CG & 100 & 0\\
     \hline
    \end{tabular}
    \caption{Algorithm comparison with 100 simulated datasets and 100 initialization of each dataset}
    \label{tab:algorithm_simulation}
\end{table}

Given that the loss values converged by the trio of algorithms are nearly indistinguishable, and considering the markedly reduced computational duration offered by \textbf{LBFGS} in comparison to its counterparts, we have elected to employ the \textbf{LBFGS} algorithm for fitting our model to the public dataset. This decision is predicated on the algorithm's enhanced efficiency in calculating our predictive matrix.

\begin{figure}[ht]
    \centering
    \includegraphics[width=0.9\textwidth]{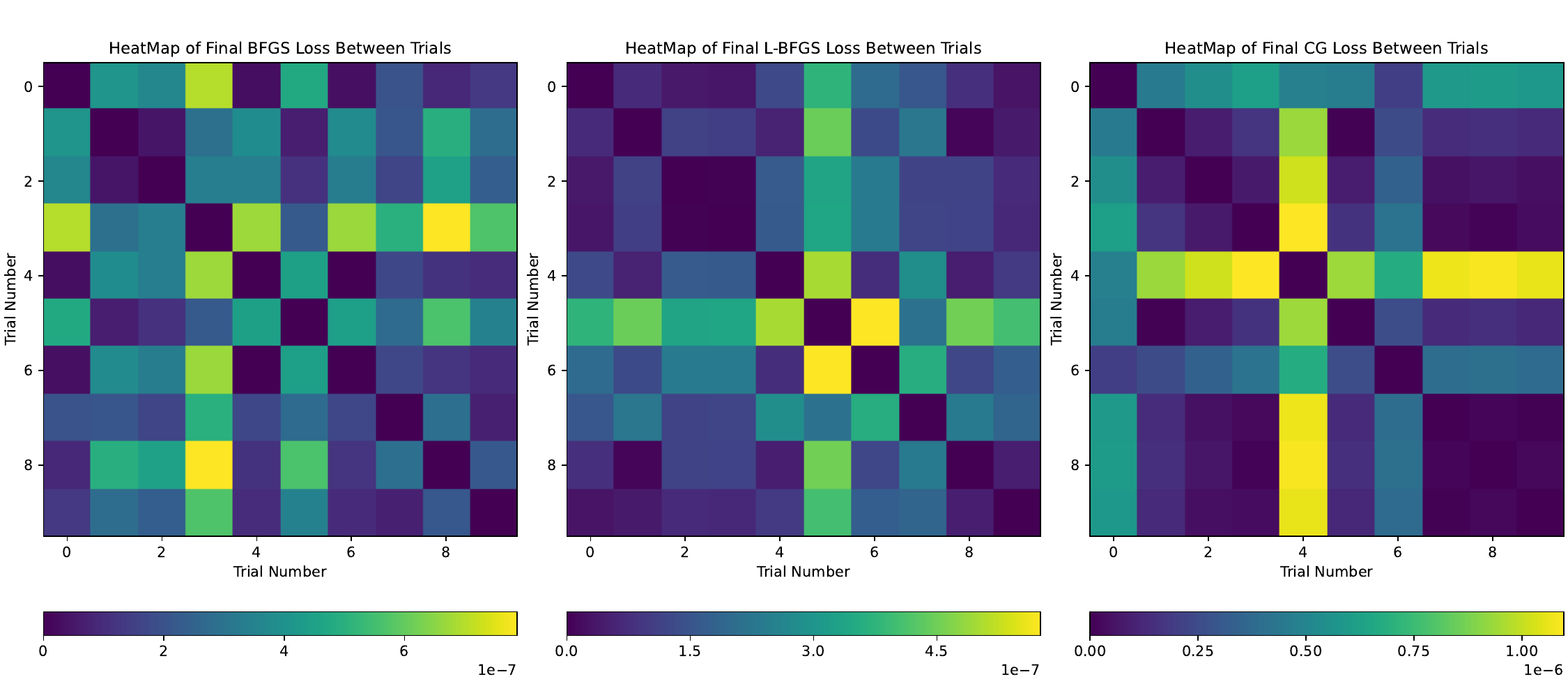}
    \caption{Heatmaps of difference of final loss value of fitted values for 45 pairs}
    \label{Heatmap1}
\end{figure}
\FloatBarrier

\subsection{Resilience of Fitted Values Estimates Under Random Initialization}
After the comparative analysis of the optimization algorithms, we investigated the stability of our model concerning varying initial conditions. For each trial, the matrices $R$ and $C$ were initialized utilizing the constant row and column means, respectively, obtained from the data preprocessing phase. Conversely, the matrices $U$ and $V$ were initialized randomly following a Standard Normal Distribution. We executed a set of 10 trials, within which we evaluated the discrepancies in the final loss values and computed the mean absolute difference for each pair of trials, culminating in a total of 45 comparative assessments. This procedure enabled us to produce heatmaps that illustrate these comparisons for each algorithm, akin to the figure presented subsequently.

\begin{figure}[ht]
    \centering
    \includegraphics[width=0.9\textwidth]{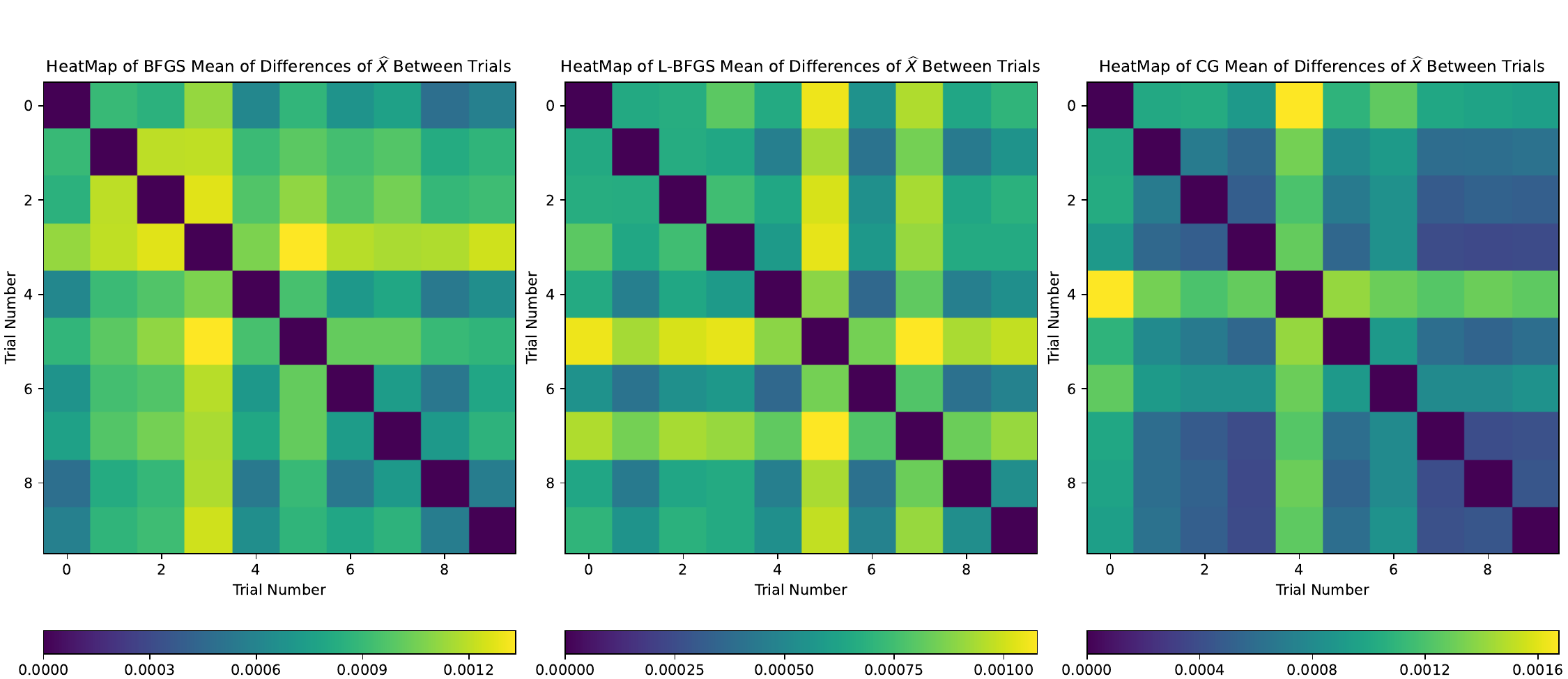}
    \caption{Heatmaps of mean of absolute difference of fitted values for 45 pairs}
    \label{Heatmap2}
\end{figure}

The outcomes depicted in figure \ref{Heatmap2} suggest that the variability in the final loss values between trial pairs is on the order of $10^{-7}$, while the mean absolute difference between trials lies in the order of $10^{-3}$. These findings positively demonstrate the model's robustness against random initializations, affirming its efficacy in reliably reconstructing data with a high degree of stability across multiple iterations.

\subsection{Effect of misspecification of model rank}
In our simulation study, we further explored the impact of the selected model rank on the loss value. Our findings indicate that when the true rank of $U$ and $V$ is denoted as $x$, and the chosen model rank $k$ is less than this true rank ($k < x$), the resultant final loss value significantly exceeds the anticipated range, deteriorating further as $k$ decreases. Conversely, when the selected model rank $k$ meets or surpasses the true rank ($k \geq x$), the final loss values align closely with expectations. However, it is noteworthy that the optimization duration escalates with an increase in $k$.

For instance, consistent with the previous discussion, the true rank of the simulated data's $U$ and $V$ matrices is 2. We conducted simulations across different ranks $k = 1, 2, 3, 4$ with a constant $\tau = 0.1$ for average of hundreds of trials. The empirical outcomes, depicted in the subsequent table, demonstrate that the final loss values for $k = 1$ are approximately an order of magnitude greater than those obtained for $k = 2, 3, 4$. Furthermore, while the loss values for $k = 2, 3, 4$ exhibit remarkable consistency, the computational time required by the algorithm significantly increases with slight overfitting for $k = 3$ and $4$, compared to the true rank of 2. We present the results for rank from 1 to 4 in the table \ref{tab:varying_k} below.

These observations can be rationalized by acknowledging that a model rank lower than the true rank leads to an underspecified model with insufficient parameters, introducing systematic bias and manifesting as an under-fitted model with elevated loss values. In contrast, a model rank exceeding the true rank results in an overspecified scenario where the model, burdened by an excess of parameters, necessitates extended periods to minimize loss, yet ultimately converges to a loss value that mirrors the expected outcome.

\begin{table}[ht]
    \centering
    \begin{tabular}{|c|c|c|c|c|}
    \hline
    Rank ($k$) & Algorithm & Final Loss & Number of Iterations & Time of Execution (seconds) \\
    \hline
    \multirow{3}{*}{$k = 1$} & BFGS & 0.05137 & 199 & 18.747 \\
    & LBFGS & 0.05137 & 27 & 9.0607 \\
    & CG & 0.06105 & 81 & 0.197 \\
    \hline
    \multirow{3}{*}{$k = 2$} & BFGS & 0.00493 & 141 & 40.856 \\
    & LBFGS & 0.00493 & 30 & 0.075 \\
    & CG & 0.00493 & 120 & 0.219 \\
    \hline
    \multirow{3}{*}{$k = 3$} & BFGS & 0.00486 & 200 & 122.947 \\
    & LBFGS & 0.00480 & 41 & 0.115 \\
    & CG & 0.00485 & 144 & 0.247 \\
    \hline
    
    \multirow{3}{*}{$k = 4$} & BFGS & 0.00481 & 198 & 230.529 \\
    & LBFGS & 0.00466 & 38 & 0.100 \\
    & CG & 0.00468 & 288 & 0.355 \\
    \hline
    \end{tabular}
    \caption{Simulation loss results with varying rank at $\tau = 0.1$}
    \label{tab:varying_k}
\end{table}

\section{Heart rate Data Modeling}

\subsection{Dataset Overview}
The dataset for our analysis comes from a study conducted at RTI International Research Institute (RTI) in 2016 \cite{furberg_2016_53894}.  The data were subsequently hosted on the Kaggle platform.   Thirty subjects were recruited and given Fitbit Fitness Trackers, which capture 24-hour heart rates as well as other parameters.  Here we focus on the 14 subjects with reported heart rate data over multiple days.

The Fitbit monitor records heart rates every five seconds.  We segmented the 24-hour day into 288 segments of five-minute duration and used the median of all recorded heart rates within each five-minute segment for each person-day as a summary for analysis.  Segments with no available data are treated as missing.  This yields data for 14 people and a total of 334 person days.  Within these days the overall proportion of missing segment values is $29.8\%$.

Heart rates largely follow a circadian pattern with a nadir (daily minimum) early in the morning, a rapid rise after awakening, a plateau or bimodal pattern in mid-day, and a gradual decline in the evening.  However a wide variety of person-day deviations from this typical pattern may occur, for example, due to work schedules, differences between weekends and weekdays, and individual differences associated with lifestyle and demographics.

We first inspected the raw data and calculated some descriptive statistics of the recorded heart rates.  In particular, we used the marginal expectiles to summarize the heart rate distribution over all person days, for each time segment.  This was done for 9 different values of $\tau$ to capture a wide range of the distribution.

\begin{figure}[ht]
\centering
\includegraphics[width=0.9\textwidth]{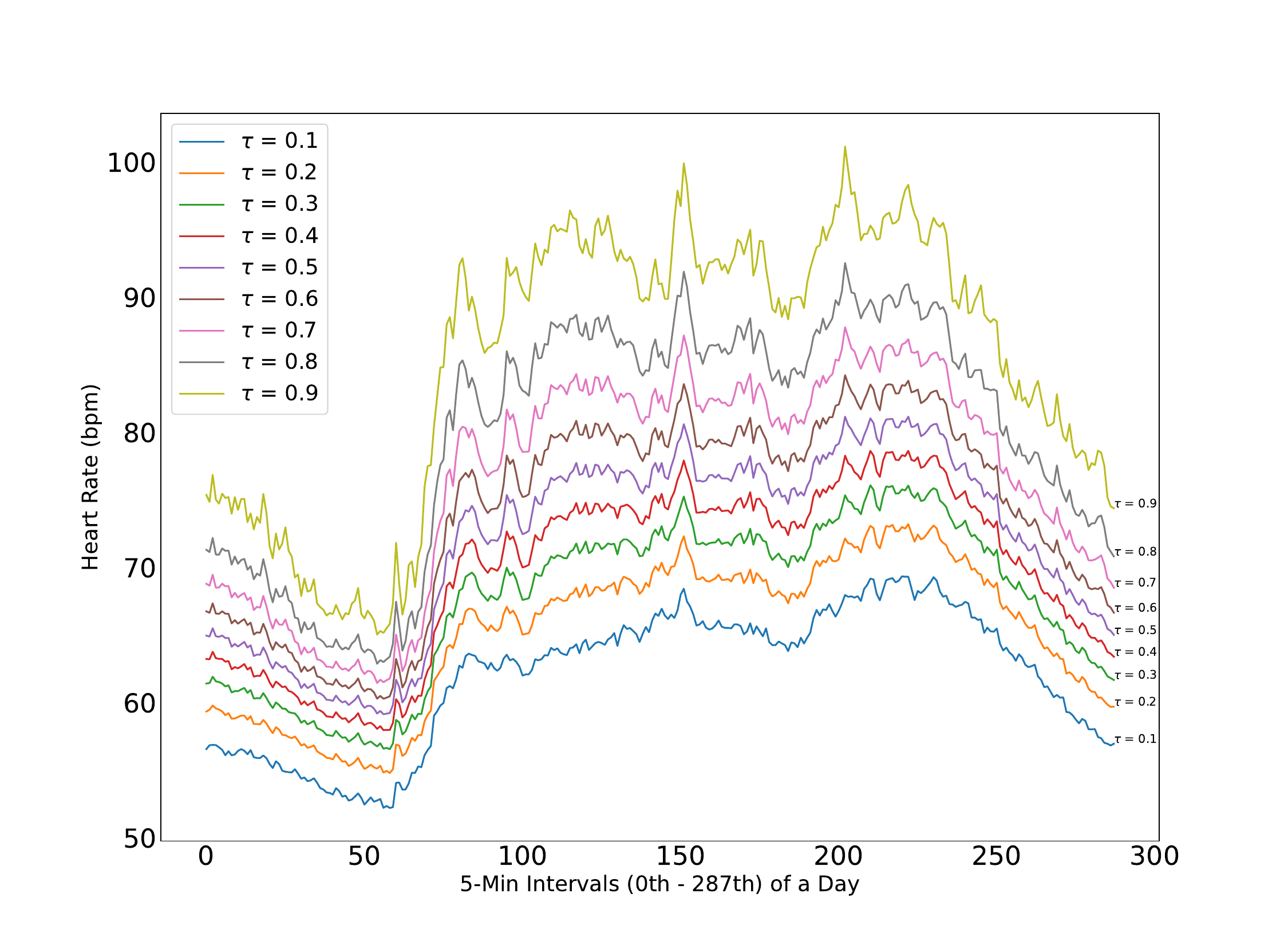} 
\caption{Marginal expectiles for each segment, overall person days} 
\label{expectile heartrate} 
\end{figure}
From figure \ref{expectile heartrate}, we observe that
the circadian heart rate pattern is broadly similar across all $\tau$ values, with the curves being translated vertically with increasing $\tau$.  However, the curves for different values of $\tau$ are notably more compressed at around the time of morning awakening and more dispersed throughout the afternoon.  Furthermore, the difference in heart rates tends to be greater between the upper expectiles compared to the lower expectiles, reflecting a right skew in the marginal heart rate distribution.

According to the simulation study, \textbf{LBFGS} is the best optimization algorithm among the three algorithms we tested. We first fit the low-rank model with $\tau = 0.5$ and the rank equals 1,  as this would produce the most stable result. We run this algorithm repeatedly and record the $R$, $C$, $U$, and $V$ for the minimum loss case, and we use the parameters of the minimum mean squared loss as the start point for other $\tau$ values.

\begin{figure}[H]
\centering
\includegraphics[width=0.85\textwidth]{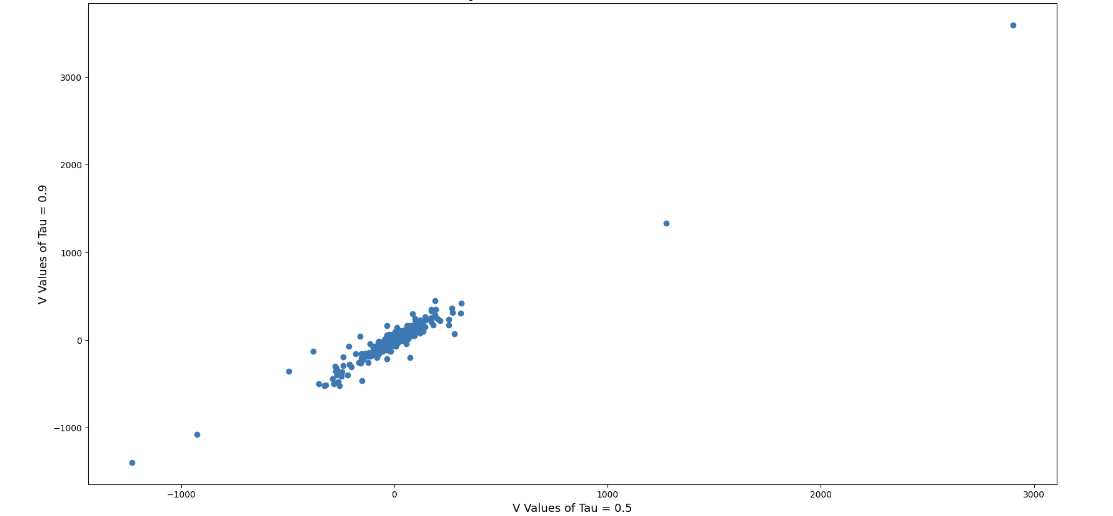} 
\caption{LBFGS's resulting normalized and oriented $V$ when $\tau = 0.5$ v.s. $\tau = 0.9$} 
\label{outlier} 
\end{figure}

One situation that arises as a result of the model is the occurrence of outliers in the predicted value of $V$, see figure \ref{outlier}.
This is due to the fact that some columns in our initial matrix have too many missing values. We remove the columns that have more than $70\%$ of the missing value in the matrix, which remains 299 columns. We re-fit the model as described above and got the expected results.

\subsection{Interpretation of fitted values}

\subsubsection{Interpretation of loss values}

Based on our model, it results in a final loss value of around 0.2658 when $\tau = 0.5$. Since the dataset has a standard deviation of $\sigma \approx 16$, we recover the average difference value between our predicted values and actual values is $\sqrt{0.2658 \times 2 \times 16^2} \approx 11.667$, giving a coefficient of determination at $\tau=0.5$ of $0.56$. In other words, our predicted heart rate value tends to lie around 11.7 bpm from the actual value.

Turning to the upper expectiles, when $\tau = 0.9$, the final loss value is around $0.1826$, and multiplying that by the variance gives us 46.75. When $\tau = 0.1$, the final loss value is around $0.094$, and multiplying that by the variance gives us 24.064. We are not certain how to interpret these two values at this point and are open to discussion. 

\subsubsection{Interpretation of parameter estimates}

\begin{figure}[ht]
\centering
\includegraphics[width=1\textwidth]{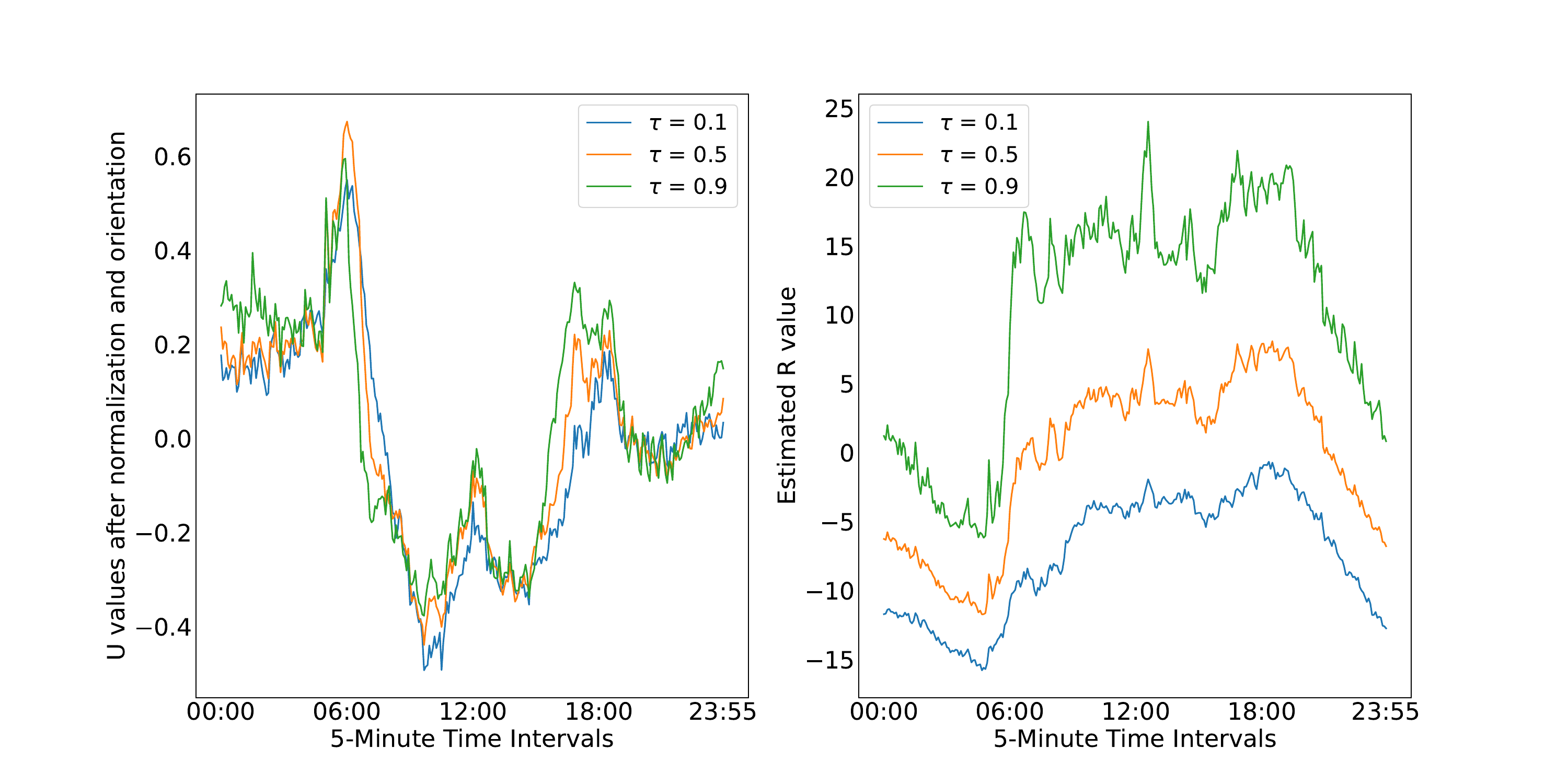} 
\caption{Result of fitted $U$ and $R$ after multiplying by standard deviation of original dataset for different expectile values} 
\label{U and R} 
\end{figure}

We multiply both the multiplicative and the additive rows by the standard deviation of the non-missing values in the original dataset to recover the fitted $U$ and $R$. The fitted values of $U$ and $R$ at $\tau=0.1, 0.5, 0.9$ are shown in figure \ref{U and R}.

The U-matrix can be interpreted as ``loadings'' in that they capture the changes across times of day that happen in coordination on any particular person x day.  Based on figure \ref{U and R}, we note that the fitted values of $U$ are quite similar regardless of $\tau$.  On the contrary, the fitted values of $R$, which represent the additive effect for each time of day, differ dramatically over $\tau$ that all three curves show some degree of volatility, which implies that the strength of the implied feature varies at different points in time. Different values of $\tau$ seem to affect the volatility of $U$. For example, when $\tau$ is 0.1 and 0.9, the fluctuations in $U$ values are more intense most of the time, while for the intermediate $\tau$ value of 0.5, the volatility seems to be slightly flatter.

The values of $R$ over time show a certain pattern that may be related to an underlying trend in the time series data. All three curves show a similar trend between approximately 08:00 and 16:00, suggesting that $R$ at different expectile levels may be influenced by the same temporal factors. In the curve for $\tau = 0.1$, there are many spikes that may indicate the occurrence of a special event, such as a violent activity.

\begin{figure}[H]
\centering
\includegraphics[width=1\textwidth]{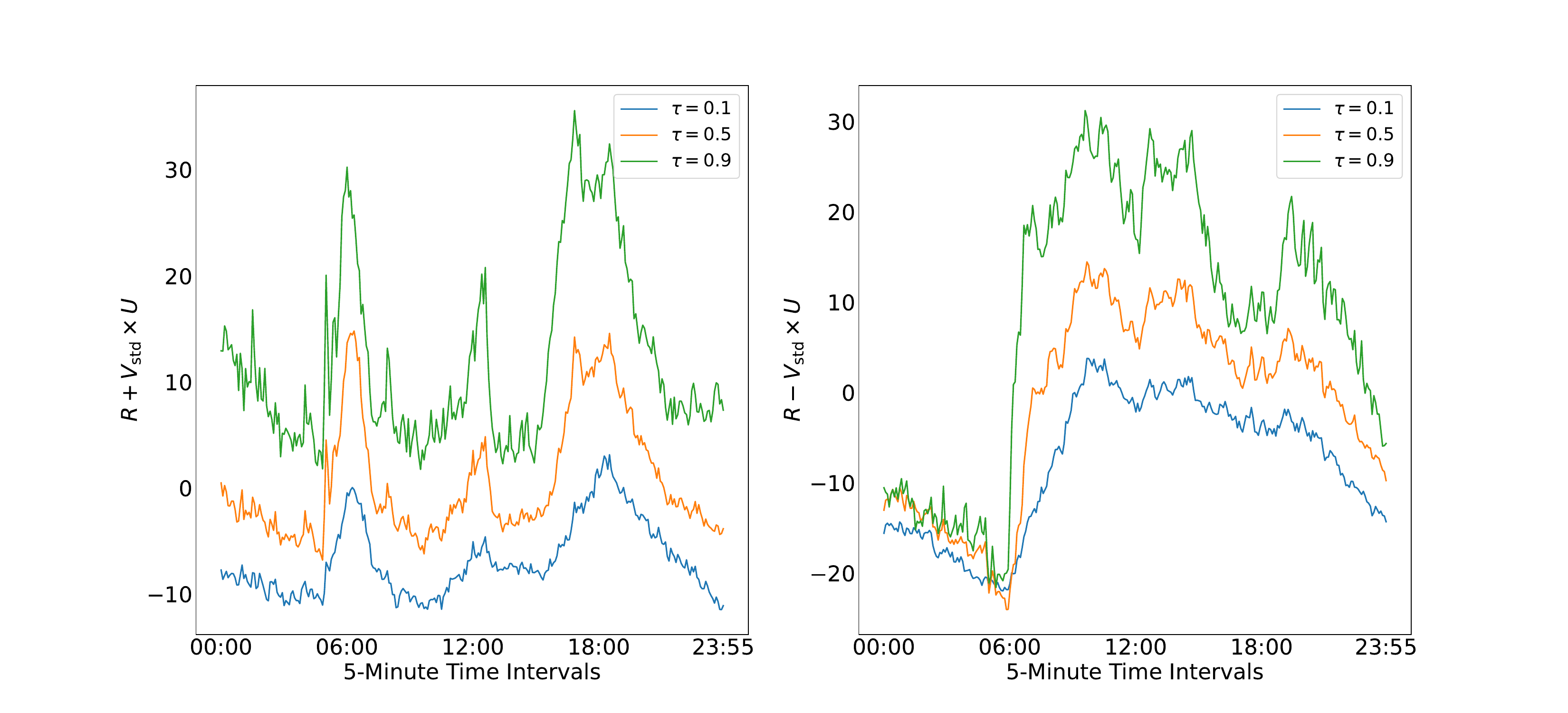} 
\caption{Plots of upper and lower values over a day for different expectile values} 
\label{R+-V_U} 
\end{figure}

The observed trends of $R+V_{\text {std }} \times U$ and $R-V_{\text {std }} \times U$ align with the role of $U V^{\top}$ in the decomposition model (see Figure \ref{R+-V_U}). All three curves $(\tau=0.1,0.5,0.9)$ exhibit periodic fluctuations over the 24 -hour time interval. The upper values of heart rates from the plot ($R+V_{\text {std }} \times U$) have trimodal distribution for different expectile values, with three peaks at around 6 a.m., 12 p.m., and 6 p.m. accordingly, which suggests a higher ceiling heart rate values usually occur during the meal times. On the other hand, the lower values of heart rates from the plot ($R-V_{\text {std }} \times U$) increase dramatically around 6 a.m. and stay high until about 6 p.m. for different expectile values, suggesting a higher floor value of the heart rates during the day, when people have activities. These patterns can aid in personalized health tracking, identifying deviations from normal heart rate trends, and detecting anomalies such as stress, fatigue, or health risks.

\begin{figure}[H]
\centering
\includegraphics[width=1\textwidth]{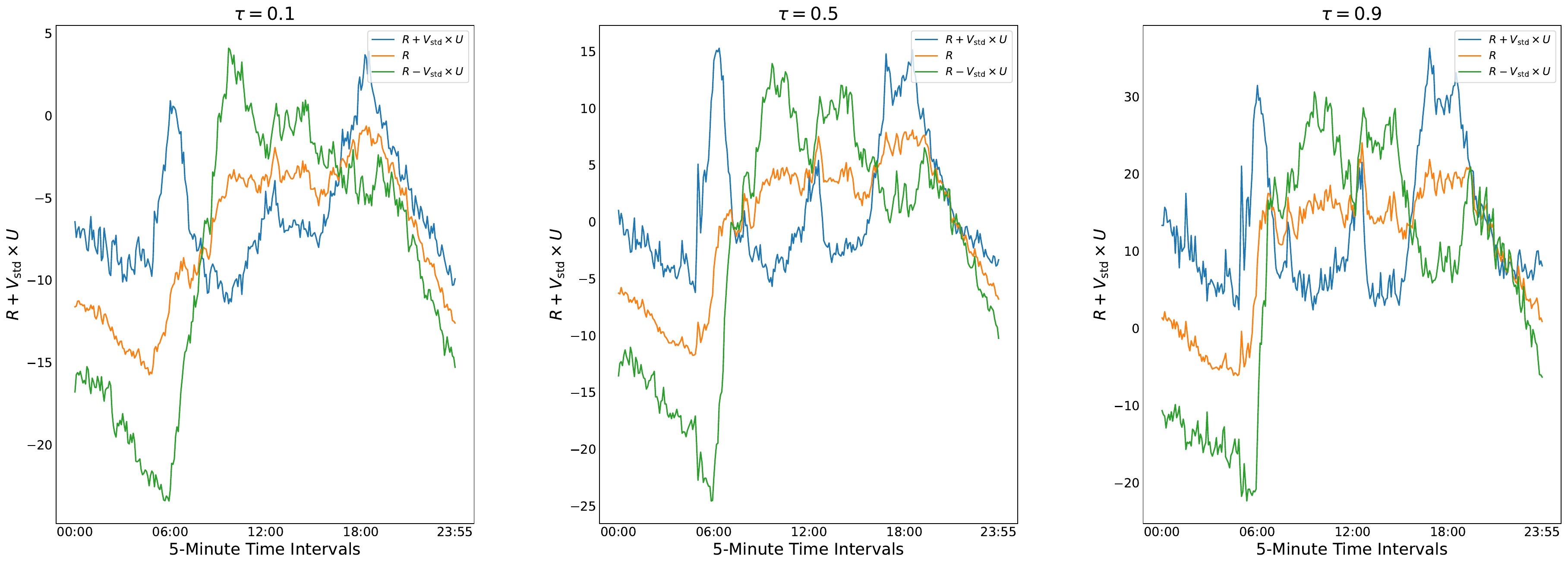} 
\caption{Plots of upper middle and lower values over a day for 3 expectile values} 
\label{R+-V_U for 3 tau} 
\end{figure}

\subsubsection{Intraclass Correlation Coefficient Analysis}
The Intraclass Correlation Coefficient (ICC) offers a means to assess the consistency of predictions across different participants and days \cite{donner1986review}. We computed ICC values for the additive component $C$ and the multiplicative component $V$ at $\tau = 0.1, 0.5,$ and $0.9$, reflecting different regions of the heart rate distribution.

\begin{figure}[H]
\centering
\includegraphics[width=0.8\textwidth]{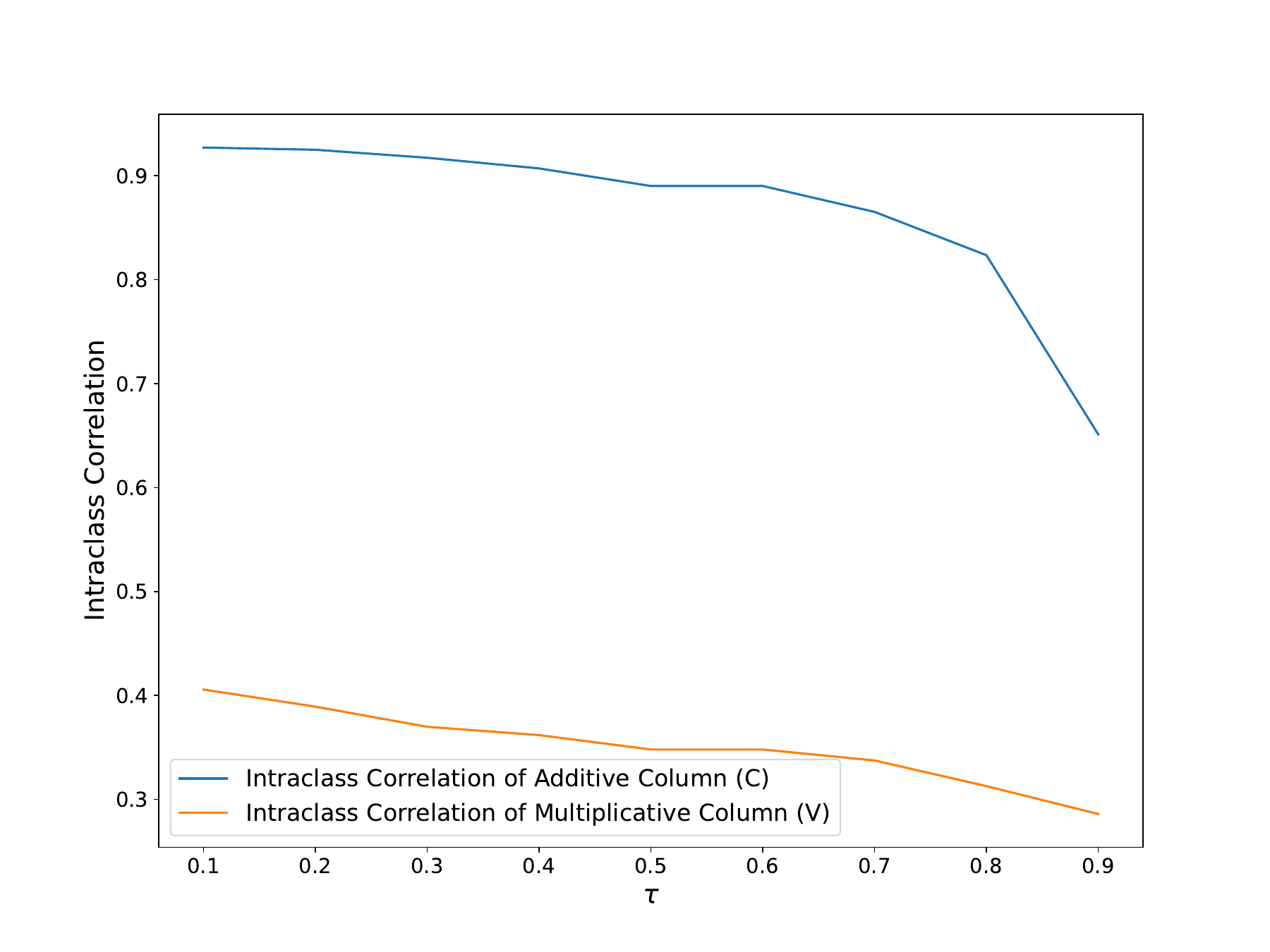} 
\caption{ICC Analysis Results}
\label{icc}
\end{figure}

For $C$, the ICC values are 0.918 at $\tau = 0.1$, 0.883 at $\tau = 0.5$, and 0.646 at $\tau = 0.9$. These higher ICCs at lower and median expectiles suggest that $C$ captures more stable individual-level features, such as overall resting or moderately active states, which vary less across participants and days. As $\tau$ increases to 0.9, the slight drop in ICC implies that the model’s estimates of $C$ are somewhat less consistent when focusing on higher heart rate tendencies, potentially reflecting behavioral fluctuations that alter the additive baseline.

In contrast, $V$ exhibits ICCs of 0.406, 0.348, and 0.285 for $\tau = 0.1, 0.5,$ and $0.9$, respectively. These relatively lower and decreasing values indicate that the component interacting with the latent factors is more sensitive to day-to-day or participant-to-participant variability in human behavior. Higher $\tau$ highlights the upper tail of the heart rate distribution, where episodic behaviors such as short bursts of vigorous activity lead to greater unpredictability in $V$. Consequently, the ICC for $V$ declines as $\tau$ increases, suggesting that modeling higher expectiles encounters greater individual-level divergence.

Overall, the gap between ICC values for $C$ and $V$ indicates that the additive component is more reliably estimated, reflecting relatively stable factors over time and across individuals, while the multiplicative component captures interactions susceptible to short-term or high-intensity behavioral differences. This aligns with the model’s aim to disentangle stable baseline effects ($C$) from dynamic factors ($V$) under different regions of the heart rate distribution \cite{donner1980estimation}.

\section{Conclusion}
This study presents a novel framework of low-rank expectile analysis, integrating additive and multiplicative effects to model diurnal heart rate patterns effectively. By defining loss function based on expectile, the approach provides a flexible tool to capture the heart rate values using a low-rank expectile representation, revealing significant insights into circadian rhythms and individual-level dynamics. The simulation results demonstrate the robustness of the model against random initialization and highlight the efficiency of the L-BFGS optimization algorithm in minimizing the expectile loss function. Through intra-class analysis, the study illustrates that the heart rate values are stable in the lower end and more variant in the upper end across different people and different days. Application to real-world heart rate data showcases the model's ability to uncover stable lower expectiles and more variable upper expectiles, emphasizing its potential in analyzing heterogeneous and noisy datasets. Future work could extend this framework to higher-dimensional low-rank matrix factorization or explore its applicability in other domains, such as blood pressure analysis or time-series anomaly detection. 

\clearpage

\bibliographystyle{alpha}
\bibliography{Low-Rank_Expectile_Representations_of_a_Data_Matrix}

\newpage
\section*{Appendices}
\appendix
\section{Derivation of Expectile Loss Gradients} \label{appendix.a}
\begin{gather*}
    \pdv{L}{R} = \pdv{(\frac{1}{N}\sum M \cdot W \cdot \widehat{E}^2)}{R} \\
    = \pdv{(\frac{1}{N}\sum(M \cdot W \cdot (X - \widehat{R}_{n \times 1}1^{\top}_{p} + 1_{n}\widehat{C}_{p \times 1}^{\top} + \widehat{U}_{n \times k}\widehat{V}^{\top}_{p \times k})^2))}{R}\\
    = -\frac{2(M \cdot W \cdot (X - \widehat{R}_{n \times 1}1^{\top}_{p} + 1_{n}\widehat{C}_{p \times 1}^{\top} + \widehat{U}_{n \times k}\widehat{V}^{\top}_{p \times k}))1^{\top}_{p} + \widehat{E} * 0} {N}\\
    = \frac{-2(M \cdot W \cdot E)1_{p}^\top}{N}
\end{gather*}

\begin{gather*}
    \pdv{L}{C} = \pdv{(\frac{1}{N}\sum(M \cdot W \cdot \widehat{E}^2)}{C} \\
    = \pdv{(\frac{1}{N}\sum(M \cdot W \cdot (X - \widehat{R}_{n \times 1}1^{\top}_{p} + 1_{n}\widehat{C}_{p \times 1}^{\top} + \widehat{U}_{n \times k}\widehat{V}^{\top}_{p \times k})^2))}{C}\\
    = \frac{-2(M \cdot W \cdot (X - \widehat{R}_{n \times 1}1^{\top}_{p} + 1_{n}\widehat{C}_{p \times 1}^{\top} + \widehat{U}_{n \times k}\widehat{V}^{\top}_{p \times k}))^\top1_{n} + \widehat{E} * 0}{N}\\
    = \frac{-2(M \cdot W \cdot \widehat{E})^\top1_{n}}{N}
\end{gather*}

\begin{gather*}
    \pdv{L}{U} = \pdv{(\frac{1}{N}\sum(M \cdot W \cdot E^2)}{U} \\
    = \pdv{(\frac{1}{N}\sum(M \cdot W \cdot (X - \widehat{R}_{n \times 1}1^{\top}_{p} + 1_{n}\widehat{C}_{p \times 1}^{\top} + \widehat{U}_{n \times k}\widehat{V}^{\top}_{p \times k})^2)}{U}\\
    = \frac{-2(M \cdot W \cdot (X - \widehat{R}_{n \times 1}1^{\top}_{p} + 1_{n}\widehat{C}_{p \times 1}^{\top} + \widehat{U}_{n \times k}\widehat{V}^{\top}_{p \times k}))V + E * 0}{N}\\
    = \frac{-2(M \cdot W \cdot E)V }{N}
\end{gather*}

\begin{gather*}
    \pdv{L}{V} = \pdv{(\frac{1}{N}\sum(M \cdot W \cdot E^2)}{V} \\
    = \pdv{(\frac{1}{N}\sum(M \cdot W \cdot (X - \widehat{R}_{n \times 1}1^{\top}_{p} + 1_{n}\widehat{C}_{p \times 1}^{\top} + \widehat{U}_{n \times k}\widehat{V}^{\top}_{p \times k})^2))}{V}\\
    = \frac{-2(M \cdot W \cdot (X - \widehat{R}_{n \times 1}1^{\top}_{p} + 1_{n}\widehat{C}_{p \times 1}^{\top} + \widehat{U}_{n \times k}\widehat{V}^{\top}_{p \times k}))^\top U + E*0}{N}\\
    = \frac{-2(M \cdot W \cdot E)^\top U}{N}
\end{gather*}

\section{Algorithms} \label{appendix.b}
\begin{algorithm}[H]
\caption{Generate Normal Simulated Data}
\begin{algorithmic}[1]
\Procedure{NormalDataGenerator}{$m, n, r\_sd, c\_sd, u\_sd, v\_sd, \sigma, na\_portion, true\_rank, seed$}
    \State Set random seed to $seed$
    \State $true\_r \gets$ Draw $m$ samples from $N(0, r\_sd)$
    \State $true\_c \gets$ Draw $n$ samples from $N(0, c\_sd)$
    \State $true\_u \gets$ Draw $m \times true\_rank$ samples from $N(0, u\_sd)$
    \State $true\_v \gets$ Draw $n \times true\_rank$ samples from $N(0, v\_sd)$
    \State $\mu\_X \gets true\_r + true\_c + true\_u \cdot true\_v^T$
    \State $error \gets \sigma \cdot m \times n \text{samples from} N(0, 1)$
    \State $X \gets \mu\_X + error$
    \State $missing \gets$ Randomly assign True/False with probability $na\_portion$
    \State Assign $NaN$ to $X$ where $missing$ is True
    \State \Return $X, true\_r, true\_c, true\_u, true\_v$
\EndProcedure
\end{algorithmic}
\end{algorithm}
In this algorithm, parameter $m$ represents the number of rows and $n$ represents the number of columns of the desired data. $r\_sd, c\_sd, u\_sd, v\_sd$ are the standard deviations for each component when drawing samples from the normal distribution, and the standard deviations have default values of 1. Each component is generated separately and the sum is the true feature matrix. $\sigma$ is the standard deviation of error we apply to a standard normal distribution and add to our true $X$. Lastly, we mark $na\_portion$ percentage of our generated data as missing values.

\begin{algorithm}[H]
\caption{Normalize Simulated Data}
\begin{algorithmic}[1]
\Procedure{GetNormalizedX}{$X$}
    \State Mask $X$ to ignore $NaN$ values
    \State $mean\_nona \gets$ Compute mean of $X$ ignoring $NaN$
    \State $std\_val \gets$ Compute standard deviation of $X$ ignoring $NaN$
    \State Normalize $X$: $X\_normalized \gets (X - mean\_nona) / std\_val$
    \State Mask $X\_normalized$ to ignore $NaN$ values
    \State Compute row and column means of $X\_normalized$ ignoring $NaN$
    \State \Return $X\_normalized, \text{row means, column means}$
\EndProcedure
\end{algorithmic}
\end{algorithm}

As illustrated above, we first calculated the mean and standard deviation of the data excluding all missing values, and applied normalization to generate the normalized data. The calculated row means and column means of the normalized data will serve as the initialization values for additive factors $R$ and $C$, as discussed in Section 2.

\begin{algorithm}[H]
\caption{Fitting of Low-Rank Model with Expectile Loss via BFGS/L-BFGS/CG}
\begin{algorithmic}[1]
\Function{TotalLossAndGradient}{$\text{params}, X, M, \tau, m, n, k$}
    \State $R, C, U, V \gets \Call{UnflattenParameters}{\text{params}, m, n, k}$
    \State $E \gets X - (R[:, \text{np.newaxis}] + C + U @ V^T)$, $E[\neg M] \gets 0$
    \State $W \gets \Call{Where}{E \geq 0, \tau, 1 - \tau}$
    \State $\text{loss} \gets \Call{Sum}{W \cdot E^2} / \Call{Sum}{M}$
    \State $R_{\text{grad}} \gets -2 \cdot \Call{Sum}{W \cdot E, \text{axis}=1} / \Call{Sum}{M}$
    \State $C_{\text{grad}} \gets -2 \cdot \Call{Sum}{W \cdot E, \text{axis}=0} / \Call{Sum}{M}$
    \State $U_{\text{grad}} \gets -2 \cdot (W \cdot E) @ V / \Call{Sum}{M}$
    \State $V_{\text{grad}} \gets -2 \cdot (W \cdot E)^T @ U / \Call{Sum}{M}$
    \State $\text{grad} \gets \Call{FlattenParameters}{R_{\text{grad}}, C_{\text{grad}}, U_{\text{grad}}, V_{\text{grad}}}$
    \State \Return $\text{loss}, \text{grad}$
\EndFunction

\Function{Optimize}{$X, \text{row\_mean}, \text{col\_mean}, k, \tau$}
    \State $m, n \gets \Call{Dimensions}{X}$, $M \gets \Call{GetObservedMask}{X}$
    \State $\text{init\_r\_param} \gets \text{row\_mean}$
    \State $\text{init\_c\_param} \gets \text{col\_mean}$
    \State $\text{init\_uv\_param} \gets \Call{RandomNormal}{0, 1, m \cdot k + n \cdot}$
    \State $\text{params} \gets \text{concatenate}(\{\text{init\_r\_param}, \text{init\_c\_param}, \text{init\_uv\_param}\})$
    \State $\text{result} \gets \Call{Minimize}{\text{TotalLossAndGradient}, \text{params}, (X, M, \tau, m, n, k), \text{'BFGS/L-BFGS/CG'}}$
    \State \Return $R, C, U, V \gets \Call{UnflattenParameters}{\text{result.x}, m, n, k}$
\EndFunction
\end{algorithmic}
\end{algorithm}
The pseudo-code above describes the model fitting procedure, where $X$ is the normalized data, $m$ and $n$ are the number of rows and columns respectively, $\tau$ is the expectile $\tau$ value, and $k$ is the picked rank value for the model. The $\Call{FLATTENPARAMETERS}$, $\Call{UNFLATTENPARAMETERS}$ are methods to manipulate parameter dimensions, and the $\Call{GETOBSERVEDMASK}$ returns where the original data holds non $NaN$ value.

\begin{algorithm}[H]
\caption{Canonicalize Parameter Estimates}
\begin{algorithmic}[1]
\Require $R, C, U, V$
\Ensure Standardized $R, C, U, V$

\For{$j = 1$ to $\text{number of columns in } V$}
    \State $R \gets R + U[:, j] \times \text{mean}(V[:, j])$
    \State $V[:, j] \gets V[:, j] - \text{mean}(V[:, j])$
\EndFor

\State $U \gets U / \text{norm}(U)$
\State $V \gets V \times \text{norm}(U)$

\State $C \gets C - \text{mean}(C)$
\State $R \gets R + \text{mean}(C)$

\State \Return $R, C, U, V$
\end{algorithmic}
\end{algorithm}
This is the procedure we used to standardize our model. Each column of $V$ has a mean of 0, $C$ has a mean of 0, and the magnitude of $U$ is 1. 

\begin{algorithm}[H]
\caption{Maintain $U$ Orientation for $k = 1$}
\begin{algorithmic}[1]
\Require $U, V$
\Ensure $U$ with consistent orientation

\If{$U[72] < 0$}
    \State $U = U \times -1$
    \State $V = V \times -1$
\EndIf

\State \Return $U, V$
\end{algorithmic}
\end{algorithm}
When the rank is 1 ($k = 1$), $U$ is of dimension $288 \times 1$. The 72nd time mark is about six a.m., which is typically the peak of $U$ in a day. If the value of $U$ at that point is negative, we will flip the curve vertically to maintain the orientation. 

\section{Code} \label{appendix.c}
All of the codes for this research project, including the processed dataset, simulation study code, and the actual project code are accessible at the Github repository: \href{https://github.com/haytham918/low-rank-expectile}{https://github.com/haytham918/low-rank-expectile}
\end{document}